\documentclass[10pt,aps,prd,showpacs,groupedaddress,notitlepage,footinbib, preprintnumbers,twocolumn]{revtex4-1}

\usepackage{braket,bm}
\usepackage[pass]{geometry} 
\usepackage{color,graphicx}
\usepackage{subfigure}
\usepackage{bm}        
\usepackage{amsmath,amsfonts,amssymb}   
\usepackage{mathrsfs,mathtools, wasysym}
\DeclareMathAlphabet{\mathpzc}{OT1}{pzc}{m}{it}

\begin{document}

\title{A new functional RG flow:~regulator-sourced 2PI versus average 1PI}

\author{Elizabeth Alexander}
\affiliation{School of Physics and Astronomy, University of Nottingham,\\ Nottingham NG7 2RD, United Kingdom}

\author{Peter Millington}
\email{p.millington@nottingham.ac.uk}
\affiliation{School of Physics and Astronomy, University of Nottingham,\\ Nottingham NG7 2RD, United Kingdom}

\author{Jordan Nursey}
\affiliation{School of Physics and Astronomy, University of Nottingham,\\ Nottingham NG7 2RD, United Kingdom}

\author{Paul M.~Saffin}
\email{paul.saffin@nottingham.ac.uk}
\affiliation{School of Physics and Astronomy, University of Nottingham,\\ Nottingham NG7 2RD, United Kingdom}

\date{August 6, 2019}

\begin{abstract}
We derive the renormalization group evolution of the quartic scalar theory with spontaneous symmetry breaking from an alternative flow equation, obtained within the externally sourced two-particle irreducible framework due to Garbrecht and Millington. In order to make a straightforward comparison with the evolution from the standard Wetterich-Morris-Ellwanger equation, we employ the Litim regulator, work to lowest order in the derivative expansion and neglect anomalous scaling. By this means, we illustrate the leading differences between analytic expressions for the resulting threshold and (non-perturbative) beta functions. In four dimensions, we find that the positions of the potential minima and the cosmological constant evolve more rapidly with scale compared to the standard approach, whereas the quartic coupling evolves more slowly, albeit by a small amount. These differences may have implications for the asymptotic safety programme, as well as our understanding of the non-perturbative scale evolution of the Standard Model Higgs sector.
\end{abstract}

\pacs{} 

\maketitle

The standard flow equation of the functional renormalization group (RG), due to Wetterich~\cite{Wetterich:1992yh}, Morris~\cite{Morris:1993qb} and Ellwanger~\cite{Ellwanger:1993mw} (and Reuter~\cite{Reuter:1996cp} in the case of gravity), can be derived from the average one-particle irreducible (1PI) effective action~\cite{Wetterich:1989xg}
\begin{equation}
\Gamma^{\rm 1PI}_{\rm av}[\phi,\mathcal{R}_k]=W[\mathcal{J}_k,\mathcal{R}_k]+\mathcal{J}_{k,x}\phi_x+\frac{1}{2}\phi_x\mathcal{R}_{k,xy}\phi_y,
\end{equation}
which depends on the RG scale $k$ through the regulator $\mathcal{R}_k$. Herein, we have used the DeWitt convention in which repeated continuous indices (here, the spacetime coordinates $x$ and $y$) are integrated over. The average 1PI effective action is defined as a `modified' Legendre transform of the (Euclidean) Schwinger function
\begin{align}
W[\mathcal{J}_k,\mathcal{R}_k]&=-\hbar \ln\int\mathcal{D}\Phi\,\exp\left[-\frac{1}{\hbar}\bigg(S[\Phi]\right.\nonumber\\&\left.\phantom{=}-\mathcal{J}_{k,x}\Phi_x-\frac{1}{2}\Phi_x\mathcal{R}_{k,xy}\Phi_y\bigg)\right].
\end{align}

Recently, however, it was shown~\cite{Alexander:2019cgw} that an alternative derivation, based on an approach to the two-particle irreducible (2PI) effective action~\cite{Cornwall:1974vz} due to Garbrecht and Millington~\cite{Garbrecht:2015cla} (see also~Ref.~\cite{Millington:2019nkw}) in which the role of the sources is fully exploited, leads to a \emph{different} flow equation. In this framework, the regulator of the RG evolution is consistently associated with the two-point source of the 2PI effective action. The regulator-sourced 2PI effective action is then defined as the following \emph{double} Legendre transform of the Schwinger function:  
\begin{align}
\Gamma^{\rm 2PI}[\phi,\Delta_k]&=W[\mathcal{J}_k,\mathcal{R}_k]+\mathcal{J}_{k,x}\phi_x\nonumber\\&\phantom{=}+\frac{1}{2}\mathcal{R}_{k,xy}\left(\phi_x\phi_y+\hbar\Delta_{k,xy}\right),
\end{align}
where $\Delta_{k,xy}$ is the source-dependent two-point function. The regulator-sourced 2PI approach differs from that of the average 1PI by the implementation of an additional Legendre transform with respect to the regulator.

The two distinct flow equations are as follows:
\begin{subequations}
\begin{align}
\label{eq:2PI}
\partial_k\Gamma^{\rm 2PI}[\phi,\Delta_k]&=+\frac{\hbar}{2}{\rm STr}\left(\mathcal{R}_k\partial_k\Delta_k\right),\\
\label{eq:av1PI}
\partial_k\Gamma^{\rm 1PI}_{\rm av}[\phi,\mathcal{R}_k]&=-\frac{\hbar}{2}{\rm STr}\left(\Delta_k\partial_k\mathcal{R}_k\right),
\end{align}
\end{subequations}
where ${\rm STr}$ indicates the supertrace over spacetime coordinates and internal indices, and $\Delta_k\mathcal{R}_k\equiv \Delta_{k,xz}\mathcal{R}_{k,zy}$. Notice that Eqs.~\eqref{eq:2PI} and \eqref{eq:av1PI} differ by a term $\partial_k\left(\mathcal{R}_k\Delta_k\right)/2$, which results from the additional Legendre transform and is, in general, non-zero (see Ref.~\cite{Alexander:2019cgw}).

In this letter, we motivate further comprehensive study of this new RG flow by identifying the key differences compared to the standard approach for the quartic scalar theory with spontaneous symmetry breaking.

We remark that we have so far employed an unusual sign convention on the regulator $\mathcal{R}_k$, which is motivated by our sign convention on the two-point source in the definition of the 2PI effective action. Otherwise, and to facilitate comparisons with the existing literature, we follow the notation of the review presented in Ref.~\cite{Berges:2000ew} as far as practicable in the remainder of this letter. (For other reviews on the functional RG, see Refs.~\cite{Pawlowski:2005xe, Gies:2006wv, Rosten:2010vm}.) Hereafter, we work in natural units, setting $\hbar =1$.

Proceeding by means of the derivative expansion, we can write the 2PI effective action as
\begin{align}
\label{eq:ansatz1}
\Gamma^{\rm 2PI}[\phi,\Delta_k]&=\int{\rm d}^dx\;\left[U_k(\rho)\right.\nonumber\\&\phantom{=}\left.+\frac{1}{2}\,Z_{k}(\rho,(\partial\phi)^2)\partial_{\mu}\phi\partial_{\mu}\phi+\mathcal{O}(\partial^4)\right],
\end{align}
where $\rho\equiv \phi^2/2$. In addition, we take the following ansatz for the form of the scale-dependent potential:
\begin{equation}
\label{eq:ansatz2}
U_k(\rho)=\frac{1}{2}g_k\left(\rho-\bar{\rho}_k\right)^2+\Lambda_k,
\end{equation}
where $g_k$, $\Lambda_k$ and $\bar{\rho}_k$ are constant with respect to $\rho$. Note that we have dropped an infinite series of terms in higher powers of $\rho-\bar{\rho}_k$. It is convenient to introduce the dimensionless parameters
\begin{subequations}
\begin{align}
\kappa_k&=\bar{Z}_kk^{2-d}\bar{\rho}_k,\\
\lambda_k&=\bar{Z}_k^{-2}k^{d-4}g_k,
\end{align}
\end{subequations}
where $\bar{Z}_k\equiv Z_k(\bar{\rho}_k,k^2)$, such that
\begin{equation}
U_k(\rho)=\frac{1}{2}k^d\lambda_k(\bar{Z}_k k^{2-d}\rho-\kappa)^2+\Lambda_k.
\end{equation}

For constant $\rho$, the regulator-sourced 2PI flow equation in Eq.~\eqref{eq:2PI} yields
\begin{align}
\label{eq:2PIflow}
\partial_t U_k(\rho)=\frac{1}{2}\int_{q}\mathcal{R}_k(q^2)\partial_t\Delta_k(\rho,q^2),
\end{align}
where $\partial_t\equiv k\partial_k$, $\int_q\equiv \int\frac{{\rm d}^dq}{(2\pi)^d}$ and
\begin{equation}
\Delta_k(\rho,q^2)=\frac{1}{Z_k(\rho,q^2) q^2-\mathcal{R}_k(q^2)+U_k'(\rho)+2\rho U''_k(\rho)}
\end{equation}
is the two-point function (again for constant $\rho$). It is convenient to re-express this in the form
\begin{equation}
\Delta_k(\rho,q^2)=\frac{1}{q^2(z_k(\rho,q^2) +r_k(q^2))+k^2\omega_k(\rho)},
\end{equation}
where we have defined
\begin{subequations}
\begin{align}
z_k(\rho,q^2)&\equiv \frac{Z_k(\rho,q^2)}{\bar{Z}_k},\\
r_k(q^2)&\equiv-\frac{\mathcal{R}_k(q^2)}{\bar{Z}_kq^2},\\
\label{eq:omega}
\omega_k(\rho)&\equiv\frac{U'_k+2\rho U''_k}{k^2\bar{Z}_k}=\lambda_k(3\bar{Z}_kk^{2-d}\rho-\kappa_k).
\end{align}
\end{subequations}

In order to make our comparison of the RG evolutions as explicit and straightforward as possible, we will work to lowest order in the derivative expansion, taking $Z_k=\bar{Z}_k$ ($z_k=1$). In addition, we will neglect all contributions from the anomalous dimension $\eta_k=-\partial_t\ln \bar{Z}_k$. We note, however, that the anomalous dimension is determined from the flow equation for the two-point function, which is itself obtained by functional differentiation of Eqs.~\eqref{eq:2PI} or~\eqref{eq:av1PI} with respect to $\phi$. It will therefore differ between the two approaches compared here.

By rewriting the integral on the right-hand side of Eq.~\eqref{eq:2PIflow} in the form
\begin{align}
\frac{1}{2}\int_q \mathcal{R}_k\left(\partial_t\Delta_k\right)&=-\frac{1}{2}\int_q\left(\partial_t \mathcal{R}_k\right)\Delta_k+\frac{1}{2}\int_q\partial_t\left(\mathcal{R}_k\Delta_k\right)\nonumber\\&\equiv 2v_dk^d\left[\ell_0^d(\omega_k)+\delta\ell_0^d(\omega_k)\right],
\end{align}
where
\begin{equation}
v_d^{-1}\equiv2^{d+1}\pi^{d/2}\Gamma(d/2),
\end{equation}
we can isolate the difference $\delta\ell_0^d(\omega_k)$ between the standard threshold function $\ell_0^d(\omega_k)$ and that which arises in the regulator-sourced 2PI flow equation. Under the coarse approximations identified above, these threshold functions are given by
\begin{subequations}
\begin{align}
\ell_0^d(\omega_k)&=+\frac{1}{4v_dk^d}\int_q\,\frac{q^2\partial_t r_k}{q^2(1+r_k)+k^2\omega_k},\\
\delta\ell_{0}^d(\omega_k)&=-\frac{1}{4v_dk^d}\int_q\,q^2\partial_t\frac{r_k}{q^2(1+r_k)+k^2\omega_k},
\end{align}
\end{subequations}
where we have suppressed the arguments on $r_k$ and $\omega_k$ for notational convenience.

We obtain the following system of flow equations for the potential, and $\kappa_k$ and $\lambda_k$, obtained respectively from Eq.~\eqref{eq:2PIflow}, and its first and second derivatives with respect to $\rho$ at $\rho=\bar{\rho}_k$ ($\omega_k=2\kappa_k\lambda_k$):
\begin{subequations}
\label{eq:flows}
\begin{align}
\partial_tU_k(\rho)&=2v_dk^d\left[\ell_0^d(\omega_k)+\delta\ell_0^d(\omega_k)\right],\\
\label{eq:kappa1}
\partial_t\kappa_k&=(2-d)\kappa_k+6v_d\left[\ell_1^d(2\kappa_k\lambda_k)+\delta\ell_{1}^d(2\kappa_k\lambda_k)\right],\\
\label{eq:lambda1}
\partial_t\lambda_k&=(d-4)\lambda_k+18v_d\lambda_k^2\left[\ell_2^d(2\kappa_k\lambda_k)+\delta\ell_{2}^d(2\kappa_k\lambda_k)\right].
\end{align}
\end{subequations}
The higher threshold functions are defined iteratively via
\begin{subequations}
\begin{align}
(\delta)\ell_{1}^d(\omega_k)&=-\partial_{\omega_k}(\delta)\ell_{0}^d(\omega_k),\\
(\delta)\ell_{2}^d(\omega_k)&=-\partial_{\omega_k}(\delta)\ell_{1}^d(\omega_k).
\end{align}
\end{subequations}
We emphasise that all contributions from the anomalous dimension have been omitted.

Making use of the Litim regulator~\cite{Litim:2002cf}
\begin{equation}
\label{eq:LitimR}
\mathcal{R}_k(q^2)=\bar{Z}_k\left(q^2-k^2\right)\Theta\left(k^2-q^2\right),
\end{equation}
the lowest-order threshold functions can be evaluated analytically, and we obtain
\begin{subequations}
\begin{align}
\ell_0^d(\omega_k)&=\frac{2}{d}\frac{1}{1+\omega_k},\\
\delta\ell_{0}^d(\omega_k)&=-\frac{2}{d+2}\frac{1}{1+\omega_k}+\frac{2}{d(d+2)}\frac{\partial_t\omega_k}{(1+\omega_k)^2}.
\end{align}
\end{subequations}
Notice that the regulator-sourced 2PI threshold function depends on $\partial_t\omega_k$. It follows from Eq.~\eqref{eq:omega} that
\begin{equation}
\partial_t\omega_k=3\lambda_k(2-d)\bar{Z}_k k^{2-d}\rho+(3\bar{Z}_k k^{2-d}\rho-\kappa_k)\partial_t\lambda_k-\lambda_k\partial_t\kappa_k.
\end{equation}
Substituting for the $\kappa_k$ and $\lambda_k$ flow equations from Eqs.~\eqref{eq:kappa1} and \eqref{eq:lambda1}, keeping only the tree-level contributions (consistent with working to first order in $\hbar$ in the flow equations), we can show that
\begin{equation}
\partial_t\omega_k=-2\omega_k.
\end{equation}
It then follows that
\begin{equation}
\delta\ell_{0}^d(\omega_k)=-\frac{2}{d+2}\frac{1}{1+\omega_k}-\frac{4}{d(d+2)}\frac{\omega_k}{(1+\omega_k)^2},
\end{equation}
and we obtain
\begin{subequations}
\begin{align}
\ell_1^d(\omega_k)&=\frac{2}{d}\frac{1}{(1+\omega_k)^2},\\
\delta\ell_{1}^d(\omega_k)&=-\frac{2(d-2)}{d(d+2)}\frac{1}{(1+\omega_k)^2}-\frac{8}{d(d+2)}\frac{\omega_k}{(1+\omega_k)^3},\\
\ell_2^d(\omega_k)&=\frac{4}{d}\frac{1}{(1+\omega_k)^3},\\
\delta\ell_{2}^d(\omega_k)&=-\frac{4(d-4)}{d(d+2)}\frac{1}{(1+\omega_k)^3}-\frac{24}{d(d+2)}\frac{\omega_k}{(1+\omega_k)^4}.
\end{align}
\end{subequations}

After substituting the threshold functions into Eq.~\eqref{eq:flows}, we arrive at the following evolution equations for the parameters $\Lambda_k$, $\kappa_k$ and $\lambda_k$:
\begin{subequations}
\label{eq:2PIflows}
\begin{align}
\label{eq:Uevonew}
\partial_t\Lambda_k&=\frac{8v_dk^d}{d(d+2)}\,\frac{1}{(1+2\kappa_k\lambda_k)^2},\\
\label{eq:betakappanew}
\partial_t\kappa_k\equiv\beta_{\kappa}&=(2-d)\kappa_k+\frac{48v_d}{d(d+2)}\,\frac{1}{(1+2\kappa_k\lambda_k)^3},\\
\partial_t\lambda_k\equiv\beta_{\lambda}&=(d-4)\lambda_k+\frac{432v_d}{d(d+2)}\,\frac{\lambda_k^2}{(1+2\kappa_k\lambda_k)^4}.
\end{align}
\end{subequations}
These results should be compared with the corresponding expressions derived from the Wetterich-Morris-Ellwanger flow equation for the same ansatz in Eqs.~\eqref{eq:ansatz1} and~\eqref{eq:ansatz2} (applied to $\Gamma^{\rm 1PI}_{\rm av}$), the same regulator in Eq.~\eqref{eq:LitimR}, and under the same coarse approximations:
\begin{subequations}
\label{eq:1PIflows}
\begin{align}
\label{eq:Uevoold}
\left.\partial_t\Lambda_k\right|_{\rm 1PI}&=\left.\frac{4v_dk^d}{d}\,\frac{1}{1+2\kappa_k\lambda_k}\right|_{\rm 1PI},\\
\label{eq:betakappaold}
\left.\partial_t\kappa_k\right|_{\rm 1PI}&=\left.(2-d)\kappa_k+\frac{12v_d}{d}\,\frac{1}{(1+2\kappa_k\lambda_k)^2}\right|_{\rm 1PI},\\
\left.\partial_t\lambda_k\right|_{\rm 1PI}&=\left.(d-4)\lambda_k+\frac{72v_d}{d}\,\frac{\lambda_k^2}{(1+2\kappa_k\lambda_k)^3}\right|_{\rm 1PI}.
\end{align}
\end{subequations}

\begin{figure}[!t]
\centering
\subfigure[\ $d=2$]{\includegraphics[width=0.44\textwidth]{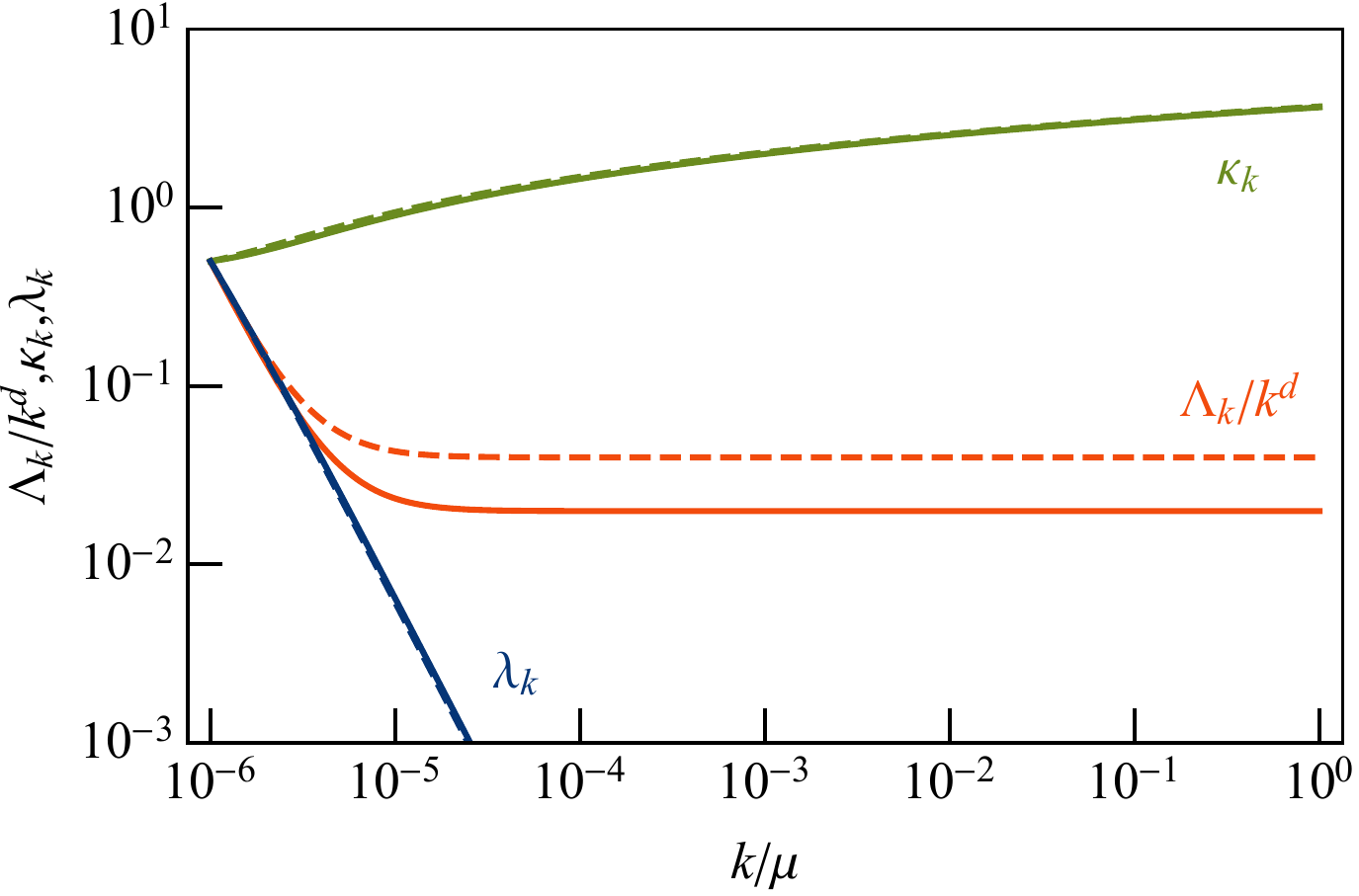}}
\subfigure[\ $d=3$]{\includegraphics[width=0.44\textwidth]{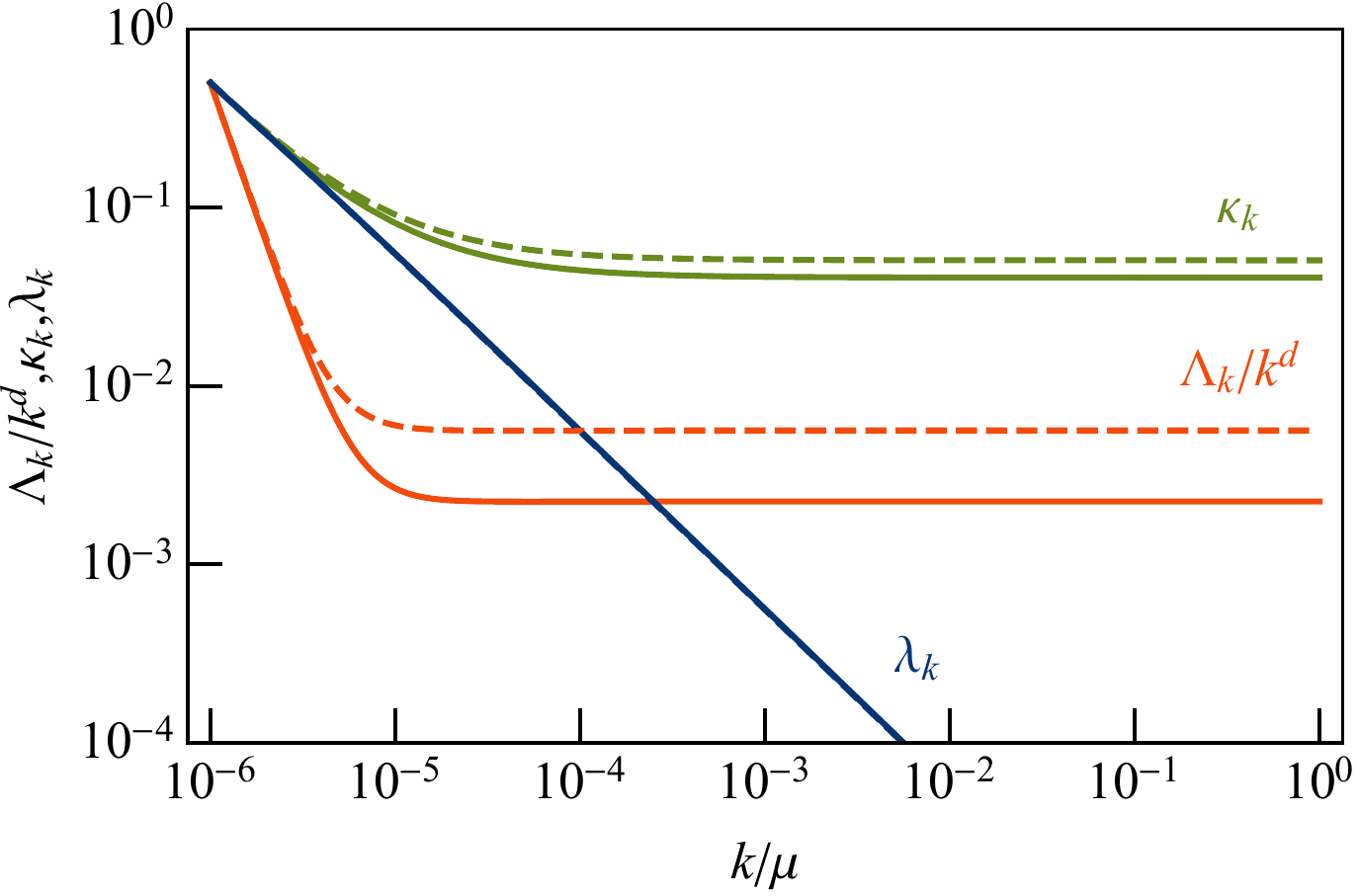}}
\subfigure[\ $d=4$]{\includegraphics[width=0.44\textwidth]{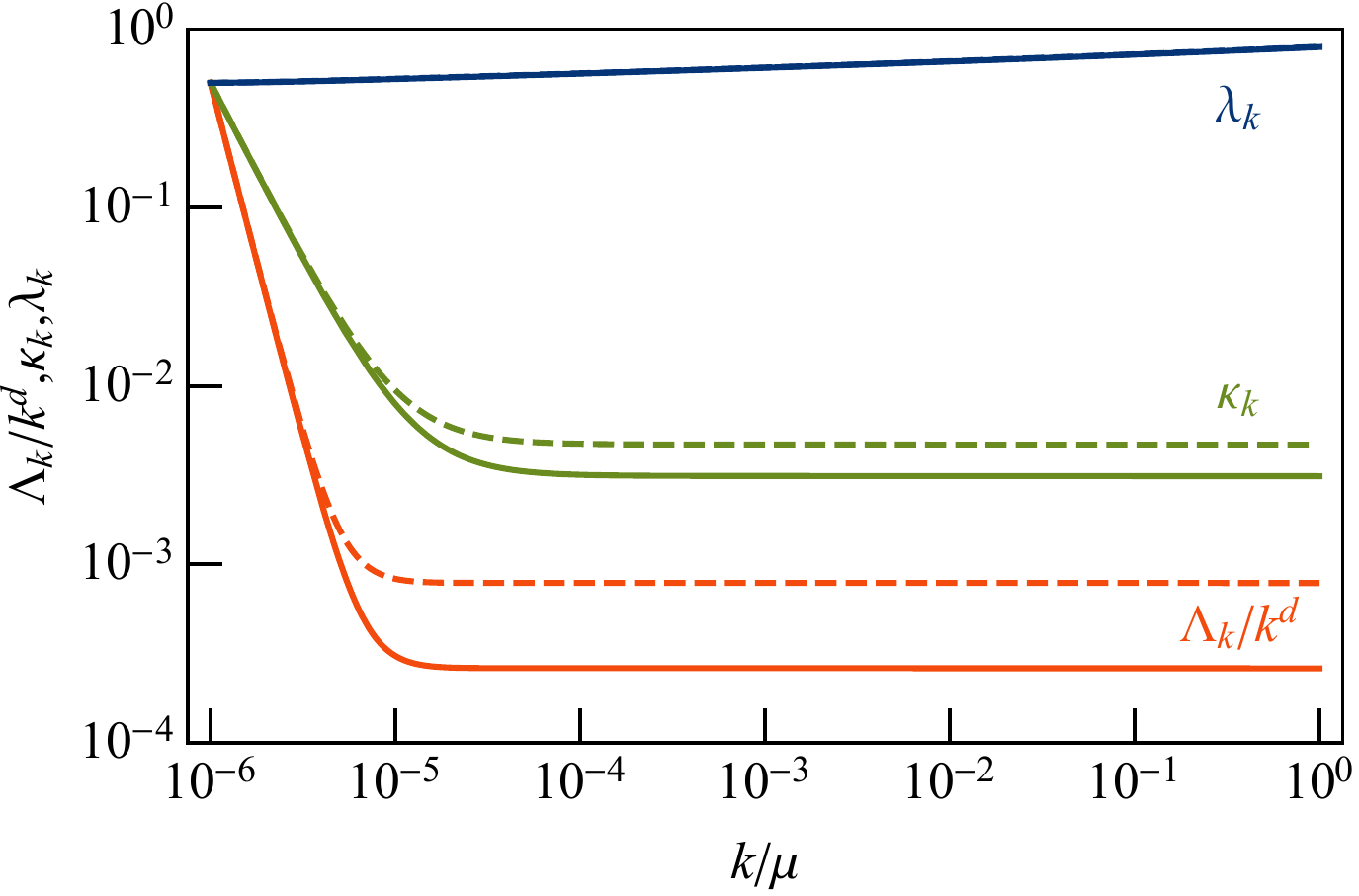}}\vspace{-0.5em}
\caption{\label{fig:flows}RG evolution of $\Lambda_k/k^d$ (orange), $\kappa_k$ (green) and $\lambda_k$ (blue) from the regulator-sourced 2PI (solid) and Wetterich-Morris-Ellwanger (dashed) flow equations to lowest order in the derivative expansion, neglecting anomalous scaling and for the Litim regulator. The boundary conditions are $\Lambda_k/k^d=\kappa_k=\lambda_k=0.5$ at $k/\mu=10^{-6}$, where $\mu$ is a mass scale.}
\end{figure}

In Fig.~\ref{fig:flows}, we plot the evolutions of the parameters $\Lambda_k/k^d$, $\kappa_k$ and $\lambda_k$ in $d=2$, $d=3$ and $d=4$, subject to the regulator-sourced 2PI and Wetterich-Morris-Ellwanger flow equations in Eqs.~\eqref{eq:2PIflows} and~\eqref{eq:1PIflows}. We see that the qualitative behaviour of the evolution is similar in each case. However, the evolution of $\kappa_k$ is faster in the regulator-sourced 2PI flow than the Wetterich-Morris-Ellwanger flow for $d=3$ and $d=4$. While not visible in the plots, the flow of $\lambda_{\kappa}$ is slower in each case, albeit by a small amount. Even so, in non-perturbative regimes, the difference in the scaling of the threshold functions may lead to more significant deviations between the two systems of flow equations.

In this letter, we have compared the regulator-sourced 2PI and average 1PI flow equations for the quartic scalar theory with spontaneous symmetry breaking. While this analysis is based on a coarse approximation, our results make clear the motivations for undertaking comprehensive comparisons to higher order in the derivative expansion and accounting fully for the anomalous scaling. The differences between these approaches may, for instance, have implications for the asymptotic safety programme~\cite{WeinbergAS} (see, e.g., Refs.~\cite{Shaposhnikov:2009pv, Eichhorn:2010tb, Dietz:2012ic, Litim:2014uca, Falls:2014tra, Bond:2016dvk, Falls:2018ylp}, and Refs.~\cite{Niedermaier:2006wt, Codello:2008vh, Reuter:2012id, Eichhorn:2018yfc} for reviews in the context of quantum gravity), as well as our understanding of the non-perturbative RG evolution of the Standard Model Higgs sector (see, e.g., Ref.~\cite{Eichhorn:2015kea}).


\begin{acknowledgements}
This work is based in part on the masters dissertations of EA and JN, supervised by PM in the School of Physics and Astronomy at the University of Nottingham. The work of PM is supported by a Leverhulme Trust Research Leadership Award, and the work of PMS is supported by the UK Science and Technology Facilities Council (STFC) under Grant No.~ST/P000703/1.
\end{acknowledgements}

\end{document}